\documentclass[preprint,showpacs,preprintnumbers,showkeys,amsmath,amssymb]{revtex4}

\usepackage{graphicx}
\usepackage{dcolumn}
\usepackage{bm}

\begin{document}


\title{Integral equation method for the electromagnetic
       wave propagation in stratified anisotropic dielectric-magnetic materials\footnote{Project supported by the National Natural
Science Foundation of China (No.~10847121,~10904036).}}

\author{SHU Wei-Xing \footnote{Corresponding author. $E$-$mail$ $address$: wxshuz@gmail.com.}}
\author{FU Na}
\author{LV Xiao-Fang}
\author{LUO Hai-Lu}
\author{WEN Shuang-Chun}
\author{FAN Dian-Yuan}

\affiliation{ Key Laboratory for Micro/Nano Optoelectronic Devices
of Ministry of Education, School of Computer and Communications,
Hunan University, Changsha 410082, China}

\begin{abstract}
We investigate the propagation of electromagnetic waves in
stratified anisotropic dielectric-magnetic materials using the
integral equation method (IEM). Based on the superposition
principle, we use Hertz vector formulations of radiated fields to
study the interaction of wave with matter. We derive in a new way
the dispersion relation, Snell's law and reflection/transmission
coefficients by self-consistent analyses. Moreover, we find two new
forms of the generalized extinction theorem. Applying the IEM, we
investigate the wave propagation through a slab and disclose the
underlying physics which are further verified by numerical
simulations. The results lead to a unified framework of the IEM for
the propagation of wave incident either from a medium or vacuum in
stratified dielectric-magnetic materials.
\end{abstract}

\keywords{integral equation, stratified dielectric-magnetic
materials, extinction theorem, metamaterials}

\pacs{41.20.Jb, 42.25.Fx, 78.20.Ci, 78.35.+C}
\maketitle

\section{Introduction}\label{Introduction}
In electrodynamics, the integral equation method (IEM) is a powerful
method for the electromagnetic wave propagation \cite{Born,Jackson}.
From a microscopic perspective, a bulk material can be regarded as a
collection of molecules (or atoms), each of which is a scatter and
emits radiation under the action of an external field. The radiated
retarded field and the exciting field interact to form the resultant
transmitted field in the material. This formulation is often
expressed as an integral equation and naturally leads to the well
known Ewald-Oseen extinction theorem which is a basis of the
scattering theory \cite{Born}. The IEM relates the macroscopic
electromagnetic responses of the material with the microscopic
features of its constituent units, thus gives much deeper physical
insight into the interaction of wave with material in contrast to
the conventional approach of Maxwell theory
\cite{Feynman1963,Wolf1972}. At the same time, the IEM does not
require boundary conditions and is advantageous in some situations,
especially when it is not sufficient to describe the interaction by
the macroscopic Maxwell equations
\cite{Wolf1972,Kong2001,Birman1972,Agarwal1973}. This approach has
been widely used to study light scattering \cite{Kong2001}, the
tip-sample interaction in near optics \cite{Girard1996,Novotny2006}
and so on. By now, the interactions of waves with semi-infinite
\cite{Reali1982}, a slab \cite{Reali1992,Lai2002} or stratified
\cite{Karam1996} isotropic dielectric media have been treated by the
IEM.

Recently, one kind of artificial anisotropic dielectric-magnetic
materials, metamaterials, has attracted considerable attention
simulated by its fascinating properties
\cite{Veselago1968,Shelby2001}. To understand the unique
characteristics of metamaterials it is necessary and appealing to
take a microscopic viewpoint. On this issue, the IEM has been
successfully implemented to predict the electric and magnetic
resonances in split-ring resonators \cite{Zhou2008}, study the
reflection of split-ring resonators \cite{Belov2006} and the imaging
process of super lens \cite{Zhou2005}, and explain the Brewster
phenomenon for TE wave associated with metamaterials
\cite{Fu2005,Shu2007b}. However, most previous work only consider
the situation of semi-infinite materials and waves incident from
vacuum. Then, how to treat the propagation of waves through
stratified materials, especially for incidence from a material
rather than vacuum? The latter problem is important because it will
result in a completely new form of the extinction theorem, as shown
in this paper.

The purpose of this paper is to use the IEM to investigate the
properties of wave propagation in stratified anisotropic
dielectric-magnetic media. Our emphasis is to disclose the
underlying physics of electromagnetic interaction with matter rather
that to obtain the well-known phenomenal results. On the basis of
the superposition principle, we use Hertz vector formulations of
radiated fields to study the microscopic mechanism of propagation.
We arrive at new derivations of the dispersion relation, Snell's law
and  reflection/transmission coefficients by self-consistent
analyses. Moreover, we find two generalized forms of the extinction
theorem. As an application, we use the IEM to study the wave
propagation through a slab and discuss the physical process of
photon tunneling, which are further verified through numerical
simulations. The results lead to a unified framework of the IEM for
the propagation of wave incident either from a medium or vacuum in
stratified dielectric-magnetic materials.

\section{the integral formulation of Hertz vectors}\label{}

In this section we take the microscopic viewpoint in which the
matter consists of molecules that react to the exciting field like
dipoles and formulate the radiated field in the form of a Green's
function integral of Hertz vectors.

\begin{figure}[t]
\centering
\includegraphics[width=8cm]{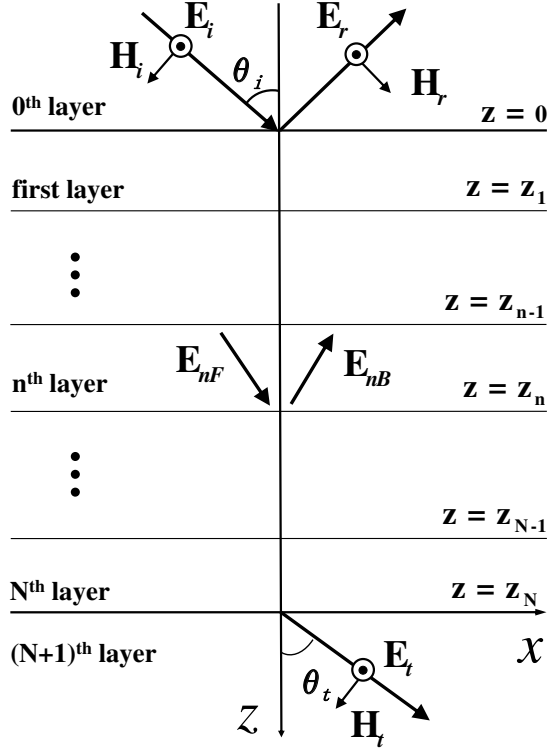}
\caption{\label{schem}Schematic diagram for the reflection and
refraction of TE waves in stratified media.}
\end{figure}

Let a monochromatic electromagnetic field of ${\bf E}_i={\bf E}_{i0}
\exp{(i{\bf k}_i\cdot {\bf r}-i\omega t)}$ and ${\bf H}_i={\bf
H}_{i0} \exp{(i{\bf k}_i\cdot {\bf r}-i\omega t)}$ incident from an
anisotropic material filling the semi-infinite space $z<0$. The
$x-z$ plane is the plane of incidence and the schematic diagram is
in Fig.~\ref{schem}. Since the material responds linearly,  all the
fields have the same dependence of $\exp{(-i\omega t)}$ which will
be omitted subsequently for simplicity. Assume the reflected fields
are ${\bf E}_r={\bf E}_{r0} \exp{(i{\bf k}_r\cdot {\bf r})}$ and
${\bf H}_r={\bf H}_{r0} \exp{(i{\bf k}_r\cdot {\bf r})}$, and the
transmitted fields are ${\bf E}_t={\bf E}_{t0} \exp{(i{\bf k}_t\cdot
{\bf r})}$ and  ${\bf H}_t={\bf H}_{t0} \exp{(i{\bf k}_t\cdot {\bf
r})}$. The permittivity and permeability tensors of the $n$-th layer
are simultaneously diagonal in the principal coordinate system,
$\boldsymbol{\varepsilon}_n= \hbox{diag}[\varepsilon_{nx},
\varepsilon_{ny}, \varepsilon_{nz}]$,
$\boldsymbol{\mu}_n=\hbox{diag}[\mu_{nx}, \mu_{ny}, \mu_{nz}]$.

Inside the layers, the incident field drive the constituent unities
to oscillate and radiate just as dipoles. The radiated electric
fields by electric and magnetic dipoles are decided by
\cite{Born,Jackson,Sein1989}
\begin{eqnarray}\label{E_dipole_field}
&&{\bf
E}_{rad}=\nabla(\nabla\cdot{\bf{\Pi}}_e)-\varepsilon_0\mu_0\frac{\partial^2{\bf{\Pi}}_e}{\partial
t^2}-\mu_0\nabla\times\frac{\partial{\bf{\Pi}}_m}{\partial t}
\end{eqnarray}
and the  magnetic fields generated by dipoles are
\begin{eqnarray}\label{H_dipole_field}
&&{\bf
H}_{rad}=\nabla(\nabla\cdot{\bf{\Pi}}_m)-\varepsilon_0\mu_0\frac{\partial^2{\bf{\Pi}}_m}{\partial
t^2}+\varepsilon_0\nabla\times\frac{\partial{\bf{\Pi}}_e}{\partial
t}.
\end{eqnarray}
Here ${\bf \Pi}_e$ and ${\bf \Pi}_m$ are the Hertz vectors, which
can be expressed as integrations of the Green's function,
\begin{eqnarray}\label{Pi_e}
&&{\bf \Pi}_e({\bf r})=\int_{V'} \frac{{\bf P}({\bf
r'})}{\varepsilon_0}G({\bf r}-{\bf r}')\hbox{d}{\bf r}',\\
&&{\bf \Pi}_m({\bf r})=\int_{V'} {\bf M}({\bf r'})G({\bf r}-{\bf
r}')\hbox{d}{\bf r}'.\label{Pi_m}
\end{eqnarray}
${\bf P}$ is the dipole moment density of electric dipoles and ${\bf
M}$ is that of magnetic dipoles, which are related to the internal
fields as ${\bf P}_{n}=\varepsilon_0\boldsymbol{\chi}_{en}\cdot{\bf
E}_{n}$, ${\bf M}_{n}=\boldsymbol{\chi}_{mn}\cdot{\bf H}_{n}$, where
the electric susceptibility
$\boldsymbol{\chi}_{en}=(\boldsymbol{\varepsilon}_{n}/{\varepsilon_0})-1$
and the magnetic susceptibility
$\boldsymbol{\chi}_{mn}=(\boldsymbol{\mu}_{n}/{\mu_0})-1$. Inside
the layer, the dipoles produce forward waves as well as backward
waves. So, we assume the dipole moment densities ${\bf P}$ and ${\bf
M}$ have the following forms
\begin{eqnarray}\label{E_t}
{\bf P}_{n}={\bf P}_{nF}\exp{(i{\bf k}_{nF}\cdot {\bf r})}+{\bf P}_{nB}\exp{(-i{\bf k}_{nB}\cdot {\bf r})}, \\
{\bf M}_{n}={\bf M}_{nF}\exp{(i{\bf k}_{nF}\cdot {\bf r})}+{\bf
M}_B\exp{(-i{\bf k}_{nB}\cdot {\bf r})},
\end{eqnarray}
where ${\bf k}_{nF}$ stands for the forward wave and $-{\bf k}_{nB}$
stands for the backward wave \cite{Lai2002, Reali1992}. The wave
numbers are to be decided by the dispersion relation yet to be
derived. The Green function is $G({\bf r}-{\bf r}')=\exp{(ik_i|{\bf
r}-{\bf r}'|)}/(4\pi|{\bf r}-{\bf r}'|)$. To evaluate the Hertz
vectors, we firstly represent the Green function in the Fourier
form. Then, substituting it into Eqs.~(\ref{Pi_e}) and (\ref{Pi_m})
and using the delta function definition and contour integration
method \cite{Reali1982,Lai2002}, the Hertz vectors of the $n$-th
layer can be evaluated as
\begin{widetext}
\begin{eqnarray}\label{Pi_ne_int}
{\bf \Pi}_{en}&=&\frac{{\bf
P}_{nF}}{\varepsilon_0}\{\frac{\exp{(i{\bf k}_{nF}\cdot {\bf
r})}}{k_{nF}^2-q_n^2}+\frac{\exp{[i{\bf q}_{nF}\cdot {\bf
r}+i(k_{nFz}-q_{nz})z_{n-1}]}}{2q_{nz}(q_{nz}-k_{nFz})}\nonumber\\
&&\left.+\frac{\exp{[i{\bf q}_{nB}\cdot {\bf
r}+i(k_{nFz}+q_{nz})z_n]}}{2q_{nz}(q_{nz}+k_{nFz})}\right\}
+\frac{{\bf P}_{nB}}{\varepsilon_0}\left\{\frac{\exp{(-i{\bf
k}_{nB}\cdot {\bf
r})}}{k_{nB}^2-b_n^2}\right.\nonumber\\&&\left.+\frac{\exp{[-i{\bf
b}_{nB}\cdot {\bf
r}-i(k_{nBz}+b_{nz})z_{n-1}]}}{2b_{nz}(b_{nz}+k_{nBz})}+\frac{\exp{[-i{\bf
b}_{nF}\cdot {\bf
r}-i(k_{nBz}-q_{nz})z_{n}]}}{2b_{nz}(b_{nz}-k_{nBz})}\right\}\nonumber\\
&&~~~~~~~~~~~~~~~~~~~~~~~~~~~~~~~~~~~~~~~~~~~~~~~~~~~~~~~~~~~~~~~~~~~~z_{n-1}\leq
z\leq z_n,
\end{eqnarray}
\end{widetext}
right to the material
\begin{widetext}
\begin{eqnarray}\label{Pi_ne_left_right}
{\bf \Pi}_{en}&=&\frac{{\bf
P}_{nF}}{\varepsilon_0}\frac{\exp{[i(k_{nFz}-q_{nz})z_{n-1}]}-\exp{[i(k_{nFz}-q_{nz})z_{n}]}}
{2q_{nz}(q_{nz}-k_{nFz})}\exp{(i{\bf q}_{nF}\cdot {\bf
r})}\nonumber\\
&&+\frac{{\bf
P}_{nB}}{\varepsilon_0}\frac{\exp{[-i(k_{nBz}+b_{nz})z_{n-1}]}-\exp{[-i(k_{nBz}+b_{nz})z_{n}]}}
{2b_{nz}(b_{nz}+k_{nBz})}\exp{(-i{\bf b}_{nB}\cdot {\bf r})},\nonumber\\
&&~~~~~~~~~~~~~~~~~~~~~~~~~~~~~~~~~~~~~~~~~~~~~~~~~~~~~~~~~~~~~~~~~~~~z>z_{n},
\end{eqnarray}
\end{widetext}
left to the material
\begin{widetext}
\begin{eqnarray}\label{Pi_ne_right_left}
{\bf \Pi}_{en}&=&\frac{{\bf
P}_{nF}}{\varepsilon_0}\frac{\exp{[i(k_{nFz}+q_{nz})z_{n}]}-\exp{[i(k_{nFz}+q_{nz})z_{n-1}]}}
{2q_{nz}(q_{nz}+k_{nFz})}\exp{(i{\bf q}_{nB}\cdot {\bf
r})}\nonumber\\
&&+\frac{{\bf
P}_{nB}}{\varepsilon_0}\frac{\exp{[-i(k_{nBz}-b_{nz})z_n]}-\exp{[-i(k_{nBz}-b_{nz})z_{n-1}]}}
{2b_{nz}(b_{nz}-k_{nBz})}\exp{(-i{\bf b}_{nF}\cdot {\bf r})},\nonumber\\
&&~~~~~~~~~~~~~~~~~~~~~~~~~~~~~~~~~~~~~~~~~~~~~~~~~~~~~~~~~~~~~~~~~~~~
z<z_{n-1},
\end{eqnarray}
\end{widetext}
where we have used the following notations
\begin{eqnarray}
{\bf q}_{nF}=k_{nFx}\hat{{\bf x}}+q_{nz}\hat{{\bf z}}, {\bf
q}_{nB}=k_{nFx}\hat{{\bf x}}-q_{nz}\hat{{\bf z}},
q_{nz}^2=k_0^2-k_{nFx}^2;\nonumber\\
{\bf b}_{nF}=k_{nBx}\hat{{\bf x}}+b_{nz}\hat{{\bf z}}, {\bf
b}_{nB}=k_{nBx}\hat{{\bf x}}-b_{nz}\hat{{\bf z}},
b_{nz}^2=k_0^2-k_{nBx}^2.
\end{eqnarray}
Note that the term with $\exp{(*)}$ equals zero  if $(*)$ contains
$z_j\rightarrow\pm\infty$, which can be found in the calculation.
Then the Hertz vectors of arbitrary layers can be found in
Eqs.~(\ref{Pi_ne_int})-(\ref{Pi_ne_right_left}).  Analogous to ${\bf
\Pi}_{ne}$, ${\bf \Pi}_{nm}$ can be obtained by interchanging ${\bf
P}_{nF}/{\varepsilon_0}$ and ${\bf P}_{nB}/{\varepsilon_0}$ with
${\bf M}_{nF}$ and ${\bf M}_{nB}$, respectively. Inserting ${\bf
\Pi}_{ne}$ and ${\bf \Pi}_{nm}$ into Eqs.~(\ref{E_dipole_field}) and
(\ref{H_dipole_field}), the radiated fields can be calculated out.

\section{Radiated Fields and Ewald-Oseen extinction theorem}

In this section, we deduce the expressions of radiated fields
generated by constituent unities in the propagation of waves in
stratified anisotropic dielectric-magnetic media. Then we use the
superposition principle to show how all the fields are combined to
yield the reflected and transmitted fields inside the individual
layers. By a self-consistent analysis we obtain the dispersion
relation, Snell's law and Fresnel's coefficients. Moreover, we find
two generalized forms of the extinction theorem.

\subsection{Fields in the $n$-th layer}

Inside the $n$-th layer, the layer by itself and the other layers
all radiate under the action of external fields. Their radiated
fields add up to the real fields inside this layer, that is,
\begin{equation}\label{E_n}
{\bf E}_{n}={\bf E}^{left.0}_{rad}+{\bf E}^{left}_{rad}+{\bf
E}^{n}_{rad}+{\bf E}^{right}_{rad}+{\bf E}^{right.N+1}_{rad}.
\end{equation}
Here the radiated field from the left semi-infinite space
\begin{equation}\label{E^{left.0}_{rad}} {\bf
E}^{left.0}_{rad}=\left\{
\begin{array}{ccc}
\displaystyle &{\bf E}_{i0}\exp{(i{\bf k}_{i}\cdot {\bf r})}, &\hbox{if}~ z<0 ~ \hbox{is vacuum;} \\
\displaystyle &-\frac{{\bf Q}({\bf q}_{0F},{\bm P}_{0F})}
{2q_{0z}(q_{0z}-k_{0Fz})}-\frac{{\bf Q}(-{\bf b}_{0B},{\bm
P}_{0B})}{2 b_{0z}(b_{0z}+k_{0Bz})}, & \hbox{or else},
\end{array}\right.
\end{equation}
the radiated field generated by the layer itself
\begin{eqnarray}\label{E^{n}_{rad}}
{\bf E}^{n}_{rad}&=&\frac{{\bf Q}({\bf k}_{nF},{\bm
P}_{nF})}{k_{nFz}^2-q_{nz}^2}
+\frac{\exp{[i(k_{nFz}-q_{nz})z_{n-1}]}}{2q_{nz}(q_{nz}-k_{nFz})}{\bf
Q}({\bf q}_{nF},{\bm
P}_{nF})\nonumber\\
&&+\frac{\exp{[i(k_{nFz}+q_{nz})z_n]}}{2q_{nz}(q_{nz}+k_{nFz})}{\bf
Q}({\bf q}_{nB},{\bm
P}_{nF})+\frac{\exp{[-i(k_{nBz}+b_n)z_{n-1}]}}{2b_{nz}(b_{nz}+k_{nBz})}{\bf
Q}(-{\bf b}_{nB},{\bm P}_{nB})
\nonumber\\
&& +\frac{{\bf Q}({\bf k}_{nB},{\bm P}_{nB})}{k_{nBz}^2-b_{nz}^2}
+\frac{\exp{[-i(k_{nBz}-b_{nz})z_n]}}{2b_{nz}(b_{nz}-k_{nBz})}{\bf
Q}(-{\bm b}_{nF},{\bm P}_{nB}),
\end{eqnarray}
the radiated field generated by  left layers $(1,\cdots,n-1)$
\begin{eqnarray}\label{E^{left}_{rad}}
{\bf
E}^{left}_{rad}&=&\sum_{j=1}^{n-1}\frac{\exp{[i(k_{jFz}-q_{jz})z_{j-1}]}-\exp{[i(k_{jFz}-q_{jz})z_{j}]}}
{2q_{jz}(q_{jz}-k_{jFz})}{\bf Q}({\bf q}_{jF},{\bm
P}_{jF})\nonumber\\
&&+\sum_{j=1}^{n-1}\frac{\exp{[-i(k_{jBz}+b_{jz})z_{j-1}]}-\exp{[-i(k_{jBz}+b_{jz})z_{j}]}}{2
b_{jz}(k_{iz}+k_{jBz})}{\bf Q}(-{\bf b}_{jB},{\bm P}_{jB}),
\end{eqnarray}
the field radiated by  right layers $(n+1,\cdots,N)$
\begin{eqnarray}\label{E^{right}_{rad}}
{\bf
E}^{right}_{rad}&=&\sum_{j=n+1}^{N}\frac{\exp{[i(k_{jFz}+q_{jz})z_{j}]}-\exp{[i(k_{jFz}+q_{jz})z_{j-1}]}}
{2q_{jz}(k_{jFz}+q_{jz})}{\bf Q}({\bf q}_{jB},{\bm
P}_{jF})\nonumber\\
&&+\sum_{j=n+1}^{N}\frac{\exp{[-i(k_{jBz}-b_{jz})z_{j}]}-\exp{[-i(k_{jBz}-b_{jz})z_{j-1}]}}{2
b_{jz}(b_{jz}-k_{jBz})}{\bf Q}(-{\bf b}_{jF},{\bm P}_{jB}).
\end{eqnarray}
the radiated field from the right semi-infinite space
\begin{eqnarray}\label{E^{right.N+1}_{rad}}
{\bf
E}^{right.N+1}_{rad}=\frac{-\exp{[i(k_{(N+1)Fz}+q_{(N+1)z})z_{N}]}}
{2q_{(N+1)z}(k_{(N+1)Fz}+q_{(N+1)z})}{\bf Q}({\bf q}_{(N+1)B},{\bm
P}_{(N+1)F}),
\end{eqnarray}
and ${\bf Q}$ is an auxiliary function
\begin{equation}
{\bf Q}({\bf
K},{\bm P})\equiv-\frac{1}{\varepsilon_0}\left[{\bf K}\times{\bf
K}\times {\bm P}+({\bf K}^2-\varepsilon_0\mu_0\omega^2){\bm
P}+\omega\mu_0\varepsilon_0{\bf K}\times {\bm M}\right]\exp{(i{\bf
K}\cdot {\bf r})}
\end{equation}
where ${\bf M}=\boldsymbol{\chi}_m\cdot [{\bf k}\times ({\bf
P}/\varepsilon_0\boldsymbol{\chi}_e)/\boldsymbol{\mu}]/\omega$.
There are two cases for the layer.

{\bf Case 1.} The layer is an anisotropic dielectric-magnetic slab.
Using ${\bf E}_{n}=[{\bf P}_{nF}\exp{(i{\bf k}_F\cdot {\bf r})}+{\bf
P}_{nB}\exp{(-i{\bf k}_B\cdot {\bf
r})}]/(\varepsilon_0\boldsymbol{\chi}_{ne})$ and inserting
Eqs.~(\ref{E^{left.0}_{rad}}) - (\ref{E^{right.N+1}_{rad}}) into
Eq.~(\ref{E_n}), we come to the following conclusions.

(i). Checking terms associated with the incident field, we find that
${\bf q}_{1F}=-{\bf b}_{1B}={\bf q}_{0F}$, that is, $q_{1z}=b_{1z}$
and $k_{1Fx}=-k_{1Bx}=k_{ix}$ that is just the Snell's law. The
incident field is related to the radiated field as
\begin{eqnarray}\label{extinction_E_left}
&&
-\frac{\exp{[i(k_{(n-1)Fz}-q_{(n-1)z})z_{n-1}]}}{2q_{(n-1)z}(q_{(n-1)z}-k_{(n-1)Fz})}{\bf
Q}({\bf q}_{(n-1)F},{\bm
P}_{(n-1)F})\nonumber\\
&&-\frac{\exp{[-i(k_{(n-1)Bz}+q_{(n-1)z})z_{n-1}]}}{2q_{(n-1)z}(q_{(n-1)z}+k_{(n-1)Bz})}{\bf
Q}({\bf q}_{(n-1)F},{\bm
P}_{(n-1)B})\nonumber\\
&&+\frac{\exp{[i(k_{nFz}-q_{nz})z_{n-1}]}}{2q_{nz}(q_{nz}-k_{nFz})}{\bf
Q}({\bf q}_{nF},{\bm
P}_{nF})+\frac{\exp{[-i(k_{nBz}+q_{nz})z_{n-1}]}}{2q_{nz}(q_{nz}+k_{nBz})}{\bf
Q}({\bf q}_{nF},{\bm P}_{nB})=0,\nonumber\\
&&
~~~~~~~~~~~~~~~~~~~~~~~~~~~~~~~~~~~~~~~~~~~~~~~~~~~~~~~~~~~~~~~~~~~~~~~~~~~~~~~~n=1,
\cdots, N.
\end{eqnarray}
This equation is actually the \textit{forward} expression of the
Ewald-Oseen extinction theorem. It shows quantitatively that the
forward part of the radiation fields at a location in the layer
extinguishes those from the regions left to that location.

(ii). Examining terms of the phase factors $\exp{(i{\bf
q}_{n2}\cdot{\bf r})}$ and $\exp{(-i{\bf b}_{n1}\cdot{\bf r})}$ in
Eq.~(\ref{E_n}), we know that ${\bf q}_{nB}=-{\bf b}_{nF}$.  At the
same time, we have
\begin{eqnarray}\label{extinction_E_right}
&&-\frac{\exp{[i(q_{(n+1)z}+k_{(n+1)Fz})z_n)]}}{2q_{(n+1)z}(q_{(n+1)z}+k_{(n+1)Fz})}{\bf
Q}({\bf q}_{(n+1)B},{\bm
P}_{(n+1)F})\nonumber\\
&&-\frac{\exp{[-i(k_{(n+1)Bz}-q_{(n+1)z})z_n]}}{2q_{(n+1)z}(q_{(n+1)z}-k_{(n+1)Bz})}{\bf
Q}({\bf q}_{(n+1)B},{\bm P}_{(n+1)B})\nonumber\\
&&+\frac{\exp{[i(q_{nz}+k_{nFz})z_n]}}{2q_{nz}(q_{nz}+k_{nFz})}{\bf
Q}({\bf q}_{nB},{\bm
P}_{nF})+\frac{\exp{[-i(k_{nBz}-q_{nz})z_n]}}{2q_{nz}(q_{nz}-k_{nBz})}{\bf
Q}({\bf q}_{nB},{\bm P}_{nB})=0,\nonumber\\
&&
~~~~~~~~~~~~~~~~~~~~~~~~~~~~~~~~~~~~~~~~~~~~~~~~~~~~~~~~~~~~~~~~~~~~~~~~~~~~~~~~n=1,
\cdots, N.
\end{eqnarray}
Eq.~(\ref{extinction_E_right}) is the \textit{backword} form of the
Ewald-Oseen extinction theorem. It indicates the extinction of the
backward vacuum fields in the layer by those from right to the
layer. Eqs.~(\ref{extinction_E_left}) and (\ref{extinction_E_right})
are two new forms of the extinction theorem that generalize the
previous results to the case of incidence from a material rather
than vacuum \cite{Wolf1972,Karam1996}.

(iii). From terms with the phase factors $\exp{(i{\bf
k}_{nF}\cdot{\bf r})}$ or $\exp{(-i{\bf k}_{nB}\cdot{\bf r})}$ in
Eq.~(\ref{E_n}) follows the dispersion relation
\begin{equation} \label{DispersionEH}
\frac{k_{nx}^2}{\mu_{nz} \varepsilon_{ny}}+\frac{k_{nz}^2}{\mu_{nx}
\varepsilon_{ny} }=\omega^2
\end{equation}
for TE waves. So $k_{nFz}=k_{nBz}$. Eq.~(\ref{DispersionEH}) relates
all the propagation modes in different layers.

{\bf Case 2.} The layer is  vacuum. Assuming the field in the layer
${\bf E}_{n}={\bf E}_{nF}\exp{(i{\bf k}_{nF}\cdot {\bf r})}+{\bf
E}_{nB}\exp{(-i{\bf k}_{nB}\cdot {\bf r})}$, we arrive at
\begin{eqnarray}\label{extinction_E_left'}
{\bf E}_{nF}&=&-
\frac{\exp{[i(k_{(n-1)Fz}-q_{(n-1)z})z_{n-1}]}}{2q_{(n-1)z}(q_{(n-1)z}-k_{(n-1)Fz})}{\bf
Q}({\bf q}_{(n-1)F},{\bm
P}_{(n-1)F})\nonumber\\&&-\frac{\exp{[-i(k_{(n-1)Bz}+q_{(n-1)z})z_{n-1}]}}{2q_{(n-1)z}(q_{(n-1)z}+k_{(n-1)Bz})}{\bf
Q}({\bf q}_{(n-1)F},{\bm P}_{(n-1)B}),
\end{eqnarray}
and
\begin{eqnarray}\label{extinction_E_right'}
{\bf
E}_{nB}&=&-\frac{\exp{[i(q_{(n+1)z}+k_{(n+1)Fz})z_n]}}{2q_{(n+1)z}(q_{(n+1)z}+k_{(n+1)Fz})}{\bf
Q}({\bf q}_{(n+1)B},{\bm
P}_{(n+1)F})\nonumber\\&&-\frac{\exp{[-i(k_{(n+1)Bz}-q_{(n+1)z})z_n]}}{2q_{(n+1)z}(q_{(n+1)z}-k_{(n+1)Bz})}{\bf
Q}({\bf q}_{(n+1)B},{\bm P}_{(n+1)B})
\end{eqnarray}
From Eqs.~(\ref{extinction_E_left'}) and
(\ref{extinction_E_right'}), it seems that the field in the layer
was only related with the adjacent regions. However, it is only a
mathematical impression because physically all the fields from the
whole space contribute to the field in the layer, as shown in
Eq.~(\ref{E_n}) obviously.

\subsection{The Reflected Field}

In region $z<0$, the contributions of radiations from all regions
form the real fields, yielding
\begin{equation}\label{E_0}
{\bf E}_{i}+{\bf E}_{r}={\bf E}^{0}_{rad}+{\bf
E}^{right}_{rad}+{\bf E}^{right.N+1}_{rad}.
\end{equation}
The field radiated by the material in the region $z<0$ can be
obtained like Eq.~(\ref{E^{n}_{rad}})
\begin{equation}\label{E^{0}_{rad}}
{\bf E}^{0}_{rad}=\frac{{\bf Q}({\bf k}_{0F},{\bm
P}_{0F})}{k_{0Fz}^2-q_{0z}^2} +\frac{{\bf Q}({\bf q}_{0B},{\bm
P}_{0F})}{2q_{0z}(q_{0z}+k_{0Fz})} +\frac{{\bf Q}({\bf k}_{0B},{\bm
P}_{0B})}{k_{0Bz}^2-b_{0z}^2} +\frac{{\bf Q}(-{\bm b}_{0F},{\bm
P}_{0B})}{2b_{0z}(b_{0z}-k_{0Bz})}.
\end{equation}
Then  we have the extinction of \textit{backward} vacuum waves
\begin{eqnarray}\label{extinction_E_0}
\frac{{\bf Q}({\bf q}_{0B},{\bm P}_{0F})}{2q_{0z}(q_{0z}+k_{0Fz})}
+\frac{{\bf Q}({\bf q}_{0B},{\bm
P}_{0B})}{2q_{0z}(q_{0z}-k_{0Bz})}-\frac{{\bf Q}({\bf q}_{1B},{\bm
P}_{1F})}{2q_{1z}(q_{1z}+k_{1Fz})}-\frac{{\bf Q}({\bf q}_{1B},{\bm
P}_{1B})}{2q_{1z}(q_{1z}-k_{1Bz})}=0.
\end{eqnarray}
The incident and reflected fields have the following forms
\begin{eqnarray}\label{E_i_r}
{\bf E}_{i}=\frac{{\bf
P}_{0F}}{\varepsilon_0\boldsymbol{\chi}_{0e}}\exp{(i{\bf
k}_{0F}\cdot {\bf r})},~~~~{\bf E}_{r}=\frac{{\bf
P}_{0B}}{\varepsilon_0\boldsymbol{\chi}_{0e}}\exp{(-i{\bf
k}_{0B}\cdot {\bf r})},~~~~~~~~~~~ z<0.
\end{eqnarray}
By solving the above equations one can obtain the dispersion
relation like Eq.~(\ref{DispersionEH}) and the reflected field. If
the region $z<0$ is vacuum, then
\begin{eqnarray}\label{E_r_vacuum}
{\bf E}_{r}=\frac{{\bf Q}({\bf q}_{1B},{\bm
P}_{1F})}{2q_{1z}(q_{1z}+k_{1Fz})} +\frac{{\bf Q}({\bf q}_{1B},{\bm
P}_{1B})}{2q_{1z}(q_{1z}-k_{1Bz})}, ~~~~~~~~~~~~~~z<0.
\end{eqnarray}
Eq.~(\ref{E_r_vacuum}) is well known in previous work in the case of
incidence from vacuum \cite{Karam1996,Lai2002}. Differently,
Eq.~(\ref{extinction_E_0}) is a completely new form of the backward
extinction theorem because the region $z<0$ is not vacuum but a
material in which the scatters are induced to generate radiated
field. Using it one can treat the problem of electromagnetic
interaction for the incidence from either vacuum or a material.

\subsection{The Transmitted Field}

In the region $z>z_N$, the radiated fields generated by all regions
form the transmitted field. That is,
\begin{equation}\label{E_N+1}
{\bf E}_{t}={\bf E}^{left.0}_{rad}+{\bf E}^{left}_{rad}+{\bf
E}^{N+1}_{rad},
\end{equation}
where the radiated field produced by dipoles in the region $z>z_N$
is
\begin{equation}\label{E^{N+1}_{rad}}
{\bf E}^{N+1}_{rad}=\frac{{\bf Q}({\bf q}_{(N+1)F},{\bm
P}_{(N+1)F})}{k_{(N+1)Fz}^2-q_{(N+1)z}^2}
+\frac{\exp{[i(k_{(N+1)Fz}-q_{(N+1)z})z_{N}]}}{2q_{(N+1)z}(q_{(N+1)z}-k_{(N+1)Fz})}{\bf
Q}({\bf q}_{(N+1)F},{\bm P}_{(N+1)F}).
\end{equation}
Hence, one can easily obtain
\begin{eqnarray}\label{extinction_E_N+1}
&&\frac{\exp{[i(k_{(N+1)Fz}-q_{(N+1)z})z_{N}]}}{2q_{(N+1)z}(q_{(N+1)z}-k_{(N+1)Fz})}{\bf
Q}({\bf q}_{(N+1)F},{\bm P}_{(N+1)F})
\nonumber\\
&&-\frac{\exp{[i(q_{Nz}-k_{NFz})z_N]}}{2q_{Nz}(q_{Nz}-k_{NFz})}{\bf
Q}({\bf q}_{NB},{\bm
P}_{NF})-\frac{\exp{[-i(k_{NBz}+q_{Nz})z_N]}}{2q_{Nz}(q_{Nz}+k_{NBz})}{\bf
Q}({\bf q}_{NB},{\bm P}_{NB})=0\nonumber\\
\end{eqnarray}
which describes that the \textit{forward} vacuum wave generated by
the material in the right self-half space is extinguished by the
counterpart from to the region $z\leq z_N$. The transmitted field is
assumed to be
\begin{eqnarray}\label{E_t}
{\bf E}_{t}=\frac{{\bf
P}_{(N+1)F}}{\varepsilon_0\boldsymbol{\chi}_{(N+1)e}}\exp{(i{\bf
k}_{(N+1)F}\cdot {\bf r})}
\end{eqnarray}
If the region $z>z_N$ is vacuum, then
\begin{eqnarray}\label{E_t_vacuum}
{\bf E}_{t}=-\frac{\exp{[i(k_{NFz}-q_{Nz})z_{N}]}}
{2q_{Nz}(q_{Nz}-k_{NFz})}{\bf Q}({\bf q}_{NF},{\bm
P}_{NF})-\frac{\exp{[-i(k_{NBz}+q_{Nz})z_{N}]}}{2
q_{Nz}(q_{Nz}+k_{NBz})}{\bf Q}({\bf q}_{NF},{\bm P}_{NB}),
\nonumber\\~~~~~~~~~~~~~~~~~~~~~~~~~~~~~~~~~~~~~~~~~~~~~~~~~~~~~~~~~~~~~~z>0.
\end{eqnarray}
By solving the equations one can obtain the transmitted field and
the dispersion relation in the region.

\subsection{Transformation Matrix}
Actually, it is not difficult to obtain the reflected field and the
transmitted field. Using Eqs.~(\ref{extinction_E_left}),
(\ref{extinction_E_right}), (\ref{extinction_E_0}), and
(\ref{extinction_E_N+1}) (Eqs.~(\ref{extinction_E_left'}),
(\ref{extinction_E_right'}), (\ref{E_r_vacuum}), and
(\ref{E_t_vacuum}) if the associated region is vacuum), one can
derive the mutual relationship of ${\bf E}_n$ in all regions
\begin{equation}\label{E_n_n+1}
\left(
\begin{array}{c} {\bm E}_{nF}\\
{\bm E}_{nB}
\end{array}\right)
=L_{n(n+1)}\left(
\begin{array}{c} {\bm E}_{(n+1)F}\\
{\bm E}_{(n+1)B}
\end{array}\right),\,n=0,\cdots,(N+1).
\end{equation}
The transformation matrix $L_{n(n+1)}$ for TE waves is
\begin{eqnarray}\label{TransformationMatrix}
L_{n(n+1)}=\frac{1}{T_{n{(n+1)}}} \left(
\begin{array}{cc} \exp{[i(k_{(n+1)Fz}-k_{nFz})z_n]}
&R_{n{(n+1)}}\exp{[-i(k_{nFz}+k_{(n+1)Fz})z_n]}\\
R_{n{(n+1)}}\exp{[i(k_{nFz}+k_{(n+1)Fz})z_n]}
&\exp{[i(k_{nFz}-k_{(n+1)Fz})z_n]}
\end{array}\right),\nonumber\\
\end{eqnarray}
where
\begin{eqnarray}
R_{n{(n+1)}}=\frac{\mu_{(n+1)x}k_{nFz}-\mu_{nx}k_{(n+1)Fz}}{\mu_{(n+1)x}k_{nFz}+\mu_{nx}k_{(n+1)Fz}},\,
T_{n{(n+1)}}=\frac{2\mu_{(n+1)x}k_{nFz}}{\mu_{(n+1)x}k_{nFz}+\mu_{nx}k_{(n+1)Fz}}.
\end{eqnarray}
Note that ${\bm E}_{(N+1)B}=0$. Then, with Eqs.~(\ref{E_i_r}),
(\ref{E_t}) and (\ref{E_n_n+1}) we finally obtain the real reflected
and transmitted fields in individual layers.

The magnetic fields can be obtained just as the corresponding
electric fields after replacing the function ${\bf Q}({\bf K},{\bm
P})$ with ${\bf O}({\bf K},{\bm M})$,
\begin{equation}\label{O}
{\bf O}({\bf K},{\bm M})\equiv-\left({\bf K}\times{\bf K}\times {\bm
M}+{\bf K}^2{\bm M}-\varepsilon_0\mu_0\omega^2{\bm M}-\omega{\bf
K}\times {\bm P}\right)\exp{(i{\bf K}\cdot {\bf r})},
\end{equation}
where ${\bf P}=\varepsilon_0\boldsymbol{\chi}_e\cdot{\bf
E}=-\varepsilon_0\boldsymbol{\chi}_e\cdot [{\bf k}\times ({\bf
M}/\boldsymbol{\chi}_m)/\boldsymbol{\varepsilon}]/\omega$.

By now, we have derived in a new way the reflected/transmitted
fields and the reflection/transmission coefficients in stratified
media. Although involving some seemingly complicated integrals that
nevertheless have simple solutions, the IEM leads almost
miraculously to the correct results. Obviously, Eqs.~(\ref{E_n_n+1})
and (\ref{R_TE}) are in agreement with the results obtained by the
formal approach of Maxwell theory \cite{Born}.

It seems from Eqs.~(\ref{extinction_E_left}),
(\ref{extinction_E_right}), (\ref{extinction_E_0}), or
(\ref{extinction_E_N+1}) as if the extinction only occurs on the
interfaces \cite{Wolf1972}. In fact, however, the vacuum fields are
extinct at every points in all regions, as shown evidently in
Eqs.~(\ref{E_n}), (\ref{E_0}) or (\ref{E_t}). In the above analyses
of this section we do not presume any boundary condition. Instead,
one can find that the result of Eq.~(\ref{E_n_n+1}) plays the role
of boundary conditions in Maxwell theory \cite{Kong2000}.

\section{Propagation through a slab}\label{sec4}

As an application of the IEM, we discuss the case of a three-layer
configuration. For the case of $N=1$,  by Eqs.~(\ref{E_i_r}) and
(\ref{E_t}) we obtain the reflection coefficient $R_{TE}={ E}_{r0}/{
E}_{i0}$ and the transmission coefficient $T_{TE}={E}_{t0}/{
E}_{i0}$ for TE waves
\begin{equation}\label{R_TE}
R_{TE}=\frac{R_{01}+R_{12}e^{i2k_{1Fz}z_1}}{1+R_{01}R_{12}e^{i2k_{1Fz}z_1}}
,~~~T_{TE}=\frac{T_{01}T_{12}e^{ik_{1Fz}z_1}}{1+R_{01}R_{12}e^{i2k_{1Fz}z_1}}.
\end{equation}
To further demonstrate the validity of the IEM, we calculate the
propagating fields in photon tunneling and compare the results by
IEM with those by solving Maxwell equations plus boundary conditions
\cite{Born,Kong2000}.

\begin{figure}[h] \centering
\includegraphics[width=8cm]{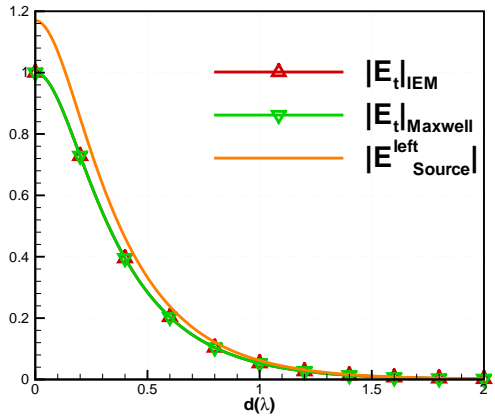}
\caption{\label{BKET}The field magnitude as a function of the width
of the slab $d$. A TE-polarized plane wave is incident on the air
gap sandwiched by two prisms with the dielectric constant $2.576$ at
the angle $\pi/4$. }
\end{figure}

\begin{figure}[h]
\includegraphics[width=4cm]{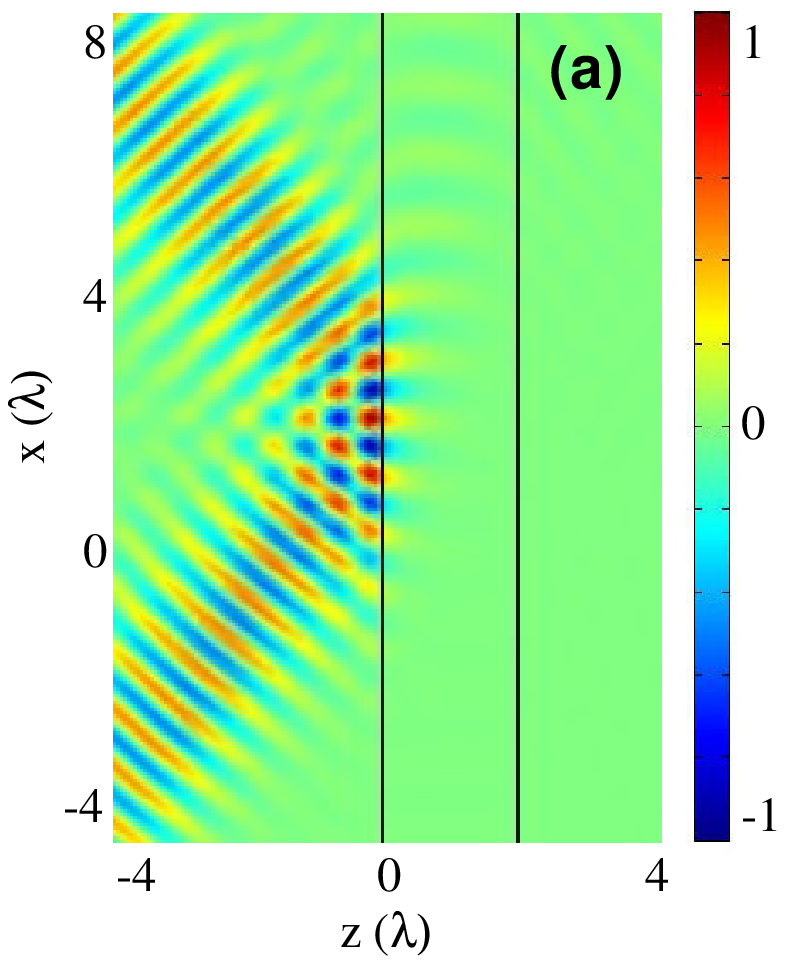}
\includegraphics[width=4cm]{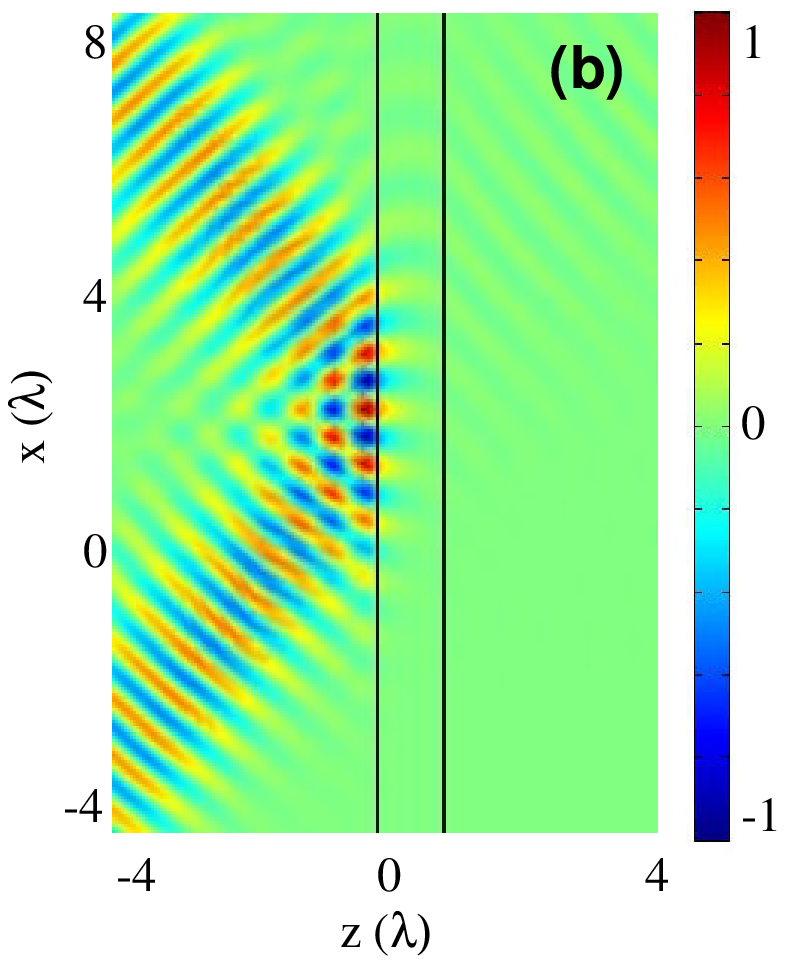}
\includegraphics[width=4cm]{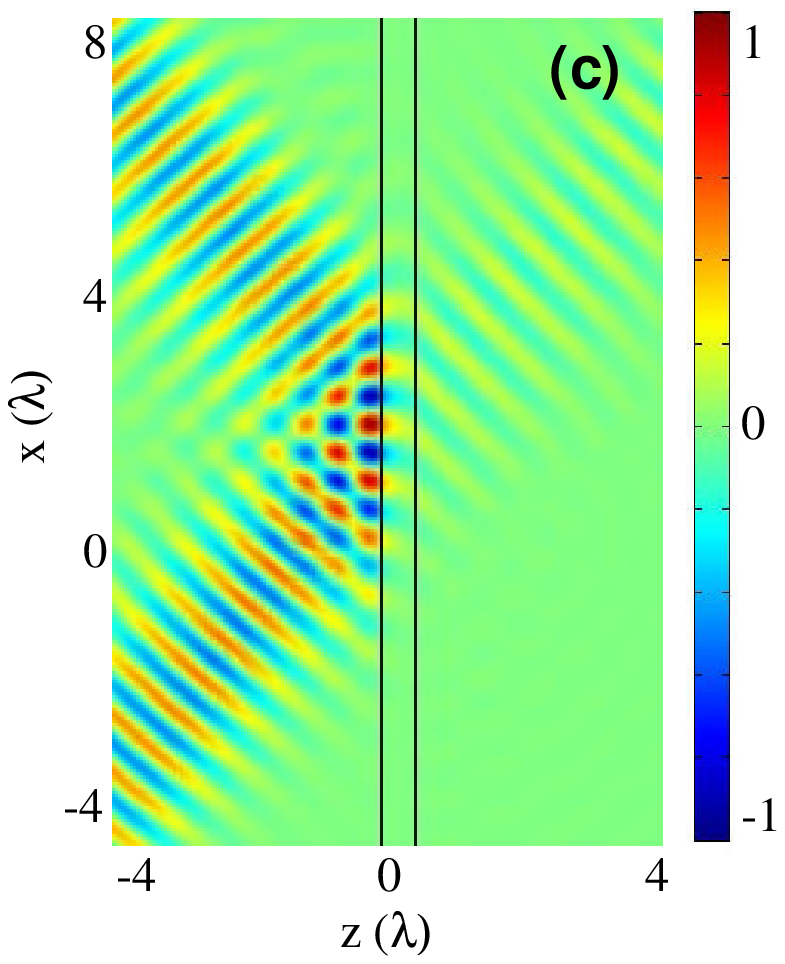}
\includegraphics[width=4cm]{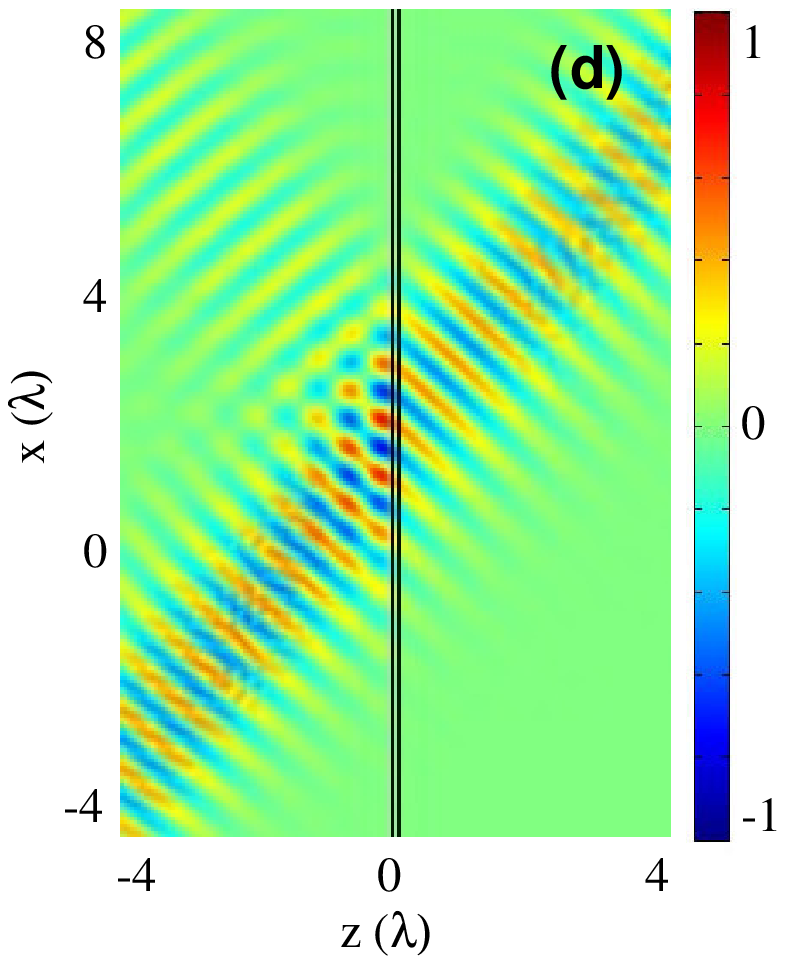}
\includegraphics[width=4cm]{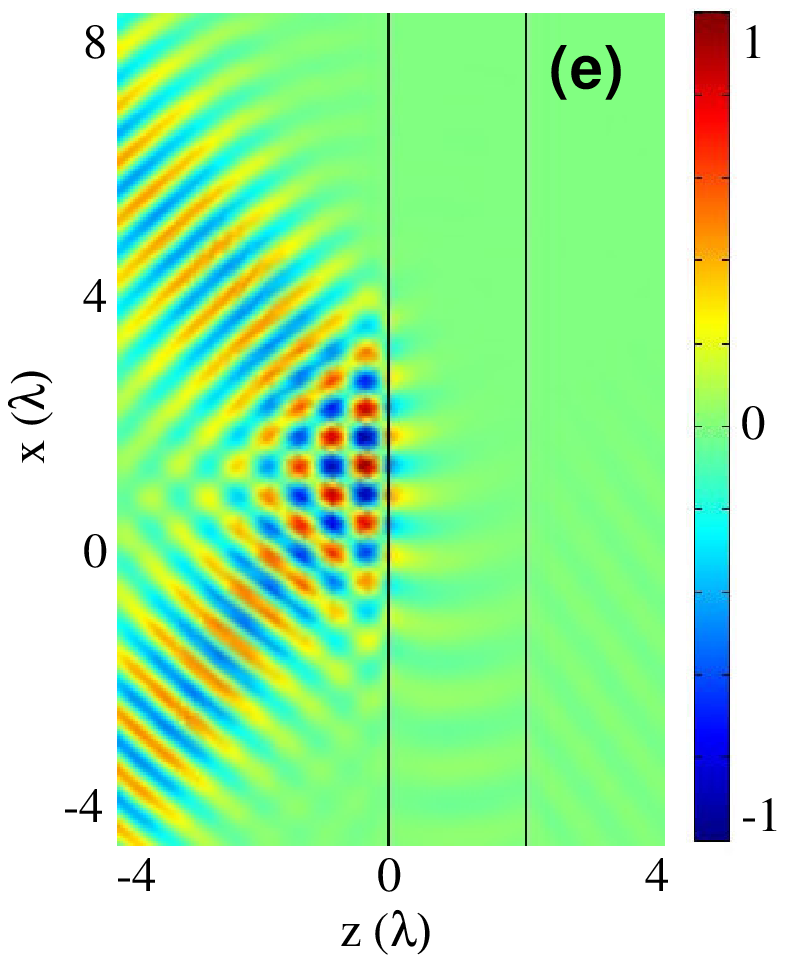}
\includegraphics[width=4cm]{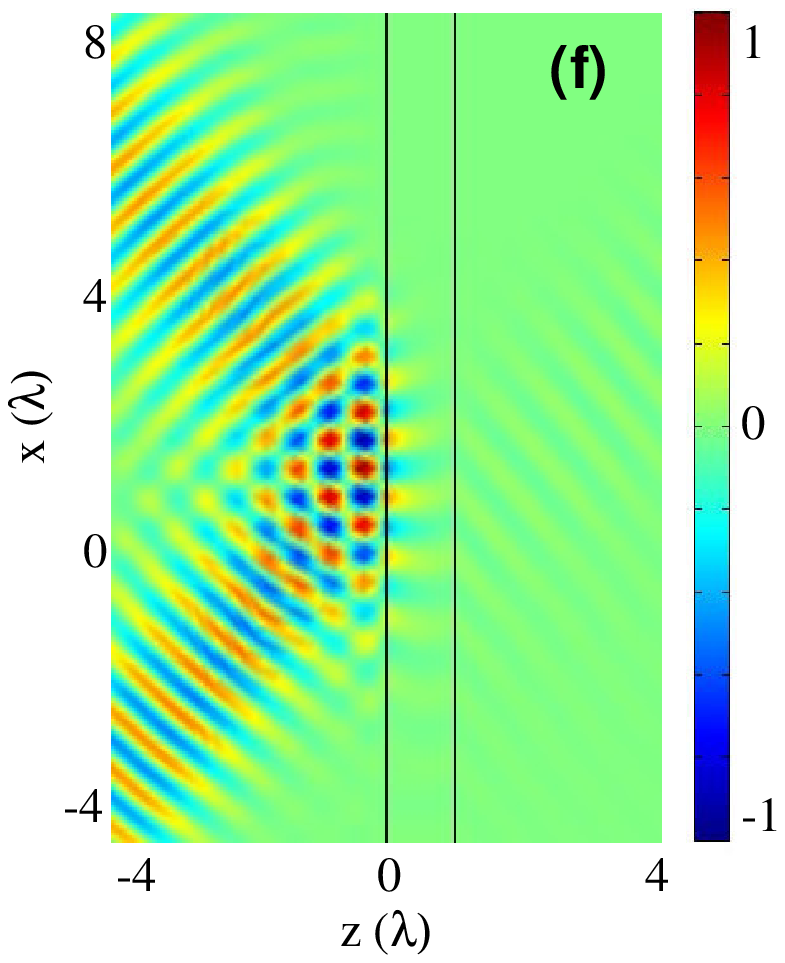}
\includegraphics[width=4cm]{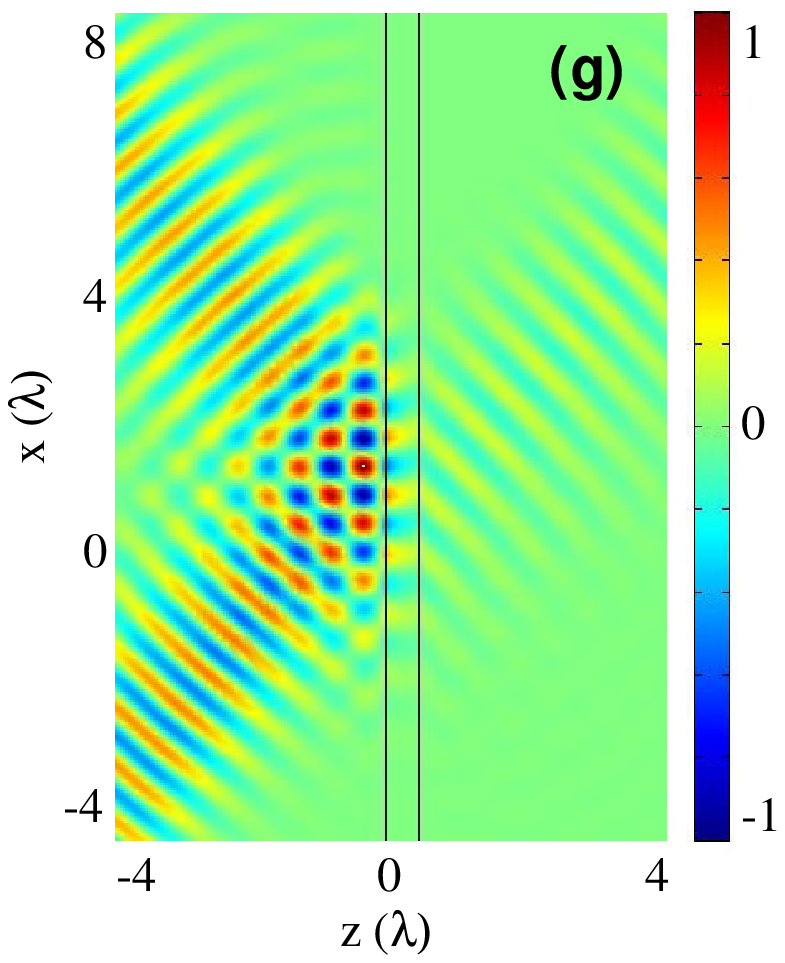}
\includegraphics[width=4cm]{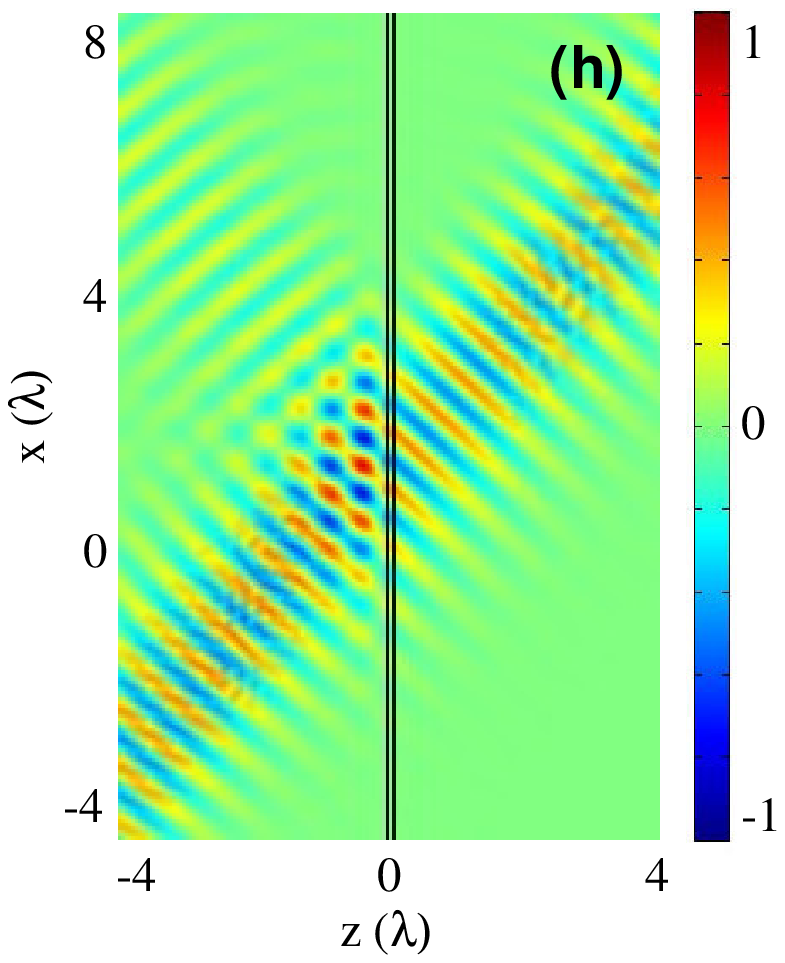}
\caption{\label{Beam}Field distributions for a Gaussian beam through
slabs with different widths $d$. (a) $d=2\lambda$, (b) $d=\lambda$,
(c) $d=0.5\lambda$, (d) $d=0.1\lambda$ for the slab with
$\varepsilon_{1y}=1$ and $\mu_{1x}=\mu_{1z}=1$; (e) $d=2\lambda$,
(f) $d=\lambda$, (g) $d=0.5\lambda$; (h) $d=0.1\lambda$ for the slab
with $\varepsilon_{1y}=-1$ and $\mu_{1x}=\mu_{1z}=-1$.  The slab is
sandwiched by two prisms with the dielectric constant $2.576$ and
the incident angle is $\pi/4$. }
\end{figure}

Consider a plane wave incident through a slab sandwiched between two
prisms at the angle larger than critical angle. We calculate the
transmitted field in the second prism by the IEM and the Maxwell
approach, denoted by $|{\bf E}_{t}|_{IEM}$ and $|{\bf
E}_{t}|_{Maxwell}$ respectively, as shown in Fig.~\ref{BKET}.
From that figure, we can see the two results overlap each other,
indicating that the IEM is consistent with Maxwell approach and thus
is valid. In order to disclose the physical mechanism of photon
tunneling, we calculate the radiated fields  $|{\bf
E}^{left}_{Source}|$ from the left to the second prism, which is the
source of the transmitted field. One can see that the amplitude of
${\bf E}^{left}_{Source}$ decreases exponentially as the width of
the slab increases. Such a trend also occurs to $|{\bf
E}_{t}|_{IEM}$. This phenomenon can be understood easily as the
transmitted field results from the radiated field induced by ${\bf
E}^{left}_{Source}$. Especially, we find that when $|{\bf
E}^{left}_{Source}|$ approaches zero, the transmitted field is also
close to zero. In other words, photon tunneling can not occur
without the radiated field ${\bf E}^{left}_{Source}$ from the left
region. Actually, detailed analyses show that the physical process
of photon tunneling is that the vacuum field ${\bf
E}^{left}_{Source}$ induce the dipoles in the second prism to
radiate two kinds of waves; one of the radiated fields extinct ${\bf
E}^{left}_{Source}$ and the other forms the final transmitted field,
leading to photon tunneling. Based on the above results, we make
numerical simulations of photon tunneling for slabs with different
widths in Fig.~\ref{Beam}. For completeness, we give results for
both positive and negative indices.
From Fig.~\ref{Beam}, one can see that the photon tunneling mostly
occurs as the slab is thinner than a wavelength and the transmitted
field increases as the slab width decreases. At the same time,
positive or negative Goos--H\"{a}nchen shifts are confirmed in
photon tunneling as observed in \cite{Kong2002}.

\section{Conclusions}\label{sec5}

In summary, we have extended the IEM to investigate the propagation
of electromagnetic waves through stratified anisotropic
dielectric-magnetic media. From a microscopic viewpoint in which the
matter consists of molecular scatters, we used the superposition
principle and the integral formulation of Hertz vectors to analyze
the interaction of field with matter. We presented new derivations
of the dispersion relation, Snell's law and the
reflection/transmission coefficients by self-consistent analyses.
Applying the IEM, we investigate the wave propagation through a slab
and discuss the physical process of photon tunneling. Verified by
numerical simulations, the results are in agreement with those
obtained by macroscopic Maxwell theory and disclose the underlying
physics of wave propagation through stratified materials, which is
the emphasis of the present paper. Moreover, we found two new forms
of the extinction theorem generalized to incidence not from vacuum
but from a material, which may enrich the theory of molecular optics
\cite{Born,Wolf1972,Karam1996}. We also demonstrated that the
extinction occurs at an arbitrary location inside the materials, not
just on the interfaces \cite{Wolf1972}. Obviously, we recovered
previous results for materials of positive refraction index
\cite{Reali1992,Lai2002,Karam1996}. More importantly, our results
lead to a unified framework of the IEM for wave propagation in
stratified dielectric-magnetic materials.

The unified IEM can be applied to other problems of propagation
through dielectric-magnetic materials, including metamaterials. This
is because the units of metamaterials, whose dimensions are much
less than the wavelength, react to exciting fields just as molecular
scatters in conventional media \cite{Pendry2006}. Especially, the
method can treat propagations of wave incident either from a medium
or vacuum, in contrast with previous work \cite{Karam1996,Lai2002}
in which only the incidence from vacuum is considered. Besides, the
same framework not only enables the handling of arbitrary shaped
objects made of continuous matter, but allows one to treat discrete
particles. The associated coupling can be included in the IEM
without any formal difficulty \cite{Girard1996,Novotny2006}. It is
hoped that the framework and the microscopic perspective of the IEM
may enable more unique properties in metamaterials to be understood
and more applications to be developed.

\end{document}